\documentclass[preprint]{aastex62}

\newcommand{\longwave}{\mbox{850 $\mu$m}}
\newcommand{\shortwave}{\mbox{450 $\mu$m}}

\graphicspath{{./}{figures/}}

\accepted{November 24, 2018}

\submitjournal{ApJ}

\shorttitle{JW 566 Flare}
\shortauthors{Mairs et al.}

\begin{document}

\title{The JCMT Transient Survey: An Extraordinary Submillimetre Flare\\ in the T Tauri Binary System JW 566}

\correspondingauthor{Steve Mairs}
\email{s.mairs@eaobservatory.org}

\author[0000-0002-6956-0730]{Steve Mairs}
\affil{East Asian Observatory,
660 N. A'ohoku Place, Hilo, HI 96720, USA}

\author[0000-0003-1618-2921]{Bhavana Lalchand}
\affiliation{Graduate Institute of Astronomy, 
National Central University, 
300 Jhongda Road, Zhongli, Taoyuan 32001, Taiwan}

\author[0000-0003-4056-9982]{Geoffrey C. Bower}
\affiliation{Academia Sinica Institute of Astronomy and Astrophysics, 645 N A'ohoku Place, Hilo, HI, 96720, USA}

\author[0000-0001-8694-4966]{Jan Forbrich}
\affiliation{Centre for Astrophysics Research, School of Physics, Astronomy and Mathematics, University of Hertfordshire,
College Lane, Hatfield AL10 9AB, UK}

\author[0000-0003-0438-8228]{Graham S. Bell}
\affil{East Asian Observatory,
660 N. A'ohoku Place, Hilo, HI 96720, USA}

\author[0000-0002-7154-6065]{Gregory J. Herczeg}
\affiliation{Kavli Institute for Astronomy and Astrophysics, 
Peking University,  Yiheyuan Lu 5,  Haidian Qu,
100871 Beijing,  People’s Republic of China}

\author[0000-0002-6773-459X]{Doug Johnstone}
\affiliation{NRC  Herzberg  Astronomy  and  Astrophysics,
5071  West Saanich Rd, 
Victoria, BC, V9E 2E7, Canada}
\affiliation{Department of Physics and Astronomy, 
University of Victoria, 
Victoria, BC, 
V8P 5C2, Canada}

\author[0000-0003-0262-272X]{Wen-Ping Chen}
\affiliation{Graduate Institute of Astronomy, National Central University, 300 Jhongda Road, Zhongli, Taoyuan 32001, Taiwan}

\author[0000-0003-3119-2087]{Jeong-Eun Lee}
\affiliation{School  of  Space  Research,  Kyung  Hee  University, 1732, Deogyeong-Daero,  Giheung-gu  Yongin-shi,  Gyunggi-do  17104, Korea}

\author[0000-0001-5397-6961]{Alvaro Hacar}
\affiliation{Leiden Observatory, Leiden University, P.O. Box 9513, 2300-RA Leiden, The Netherlands}

\begin{abstract}

The binary T Tauri system JW 566 in the Orion Molecular 
Cloud underwent an energetic, short-lived flare observed at 
submillimetre wavelengths by the SCUBA-2 instrument on 26 November 2016 (UT). 
The emission faded by nearly 50\% during the 31 minute 
integration. The simultaneous source fluxes averaged over the observation are
\mbox{$500\pm107\mathrm{\:mJy\:beam}^{-1}$} at \mbox{450 $\mu$m}
and \mbox{$466\pm47\mathrm{\:mJy\:beam}^{-1}$} at \longwave. The \mbox{850 $\mu\mathrm{m}$} 
flux corresponds to a radio luminosity of 
\mbox{$L_{\nu}=8\times10^{19}\mathrm{\:erg\:s^{-1}\:Hz^{-1}}$},
approximately one order of magnitude brighter (in 
terms of $\nu L_{\nu}$) than that of a flare of the young star 
GMR-A, detected in Orion in 2003 at 3mm. The event may be the most 
luminous known flare associated with a young 
stellar object and is also the first coronal flare discovered at sub-mm wavelengths.
The spectral index between \shortwave$\,$ and \longwave$\,$ 
of $\alpha = 0.11$ is broadly consistent
with non-thermal emission. The brightness 
temperature was in excess of $6\times10^{4}\mathrm{\:K}$. We interpret this event to be a 
magnetic reconnection that energised charged particles 
to emit gyrosynchrotron/synchrotron radiation.

\end{abstract}

\keywords{stars: formation --- surveys --- stars: variables: general --- ISM: jets and outflows}

\section{Introduction} 
\label{sec:intro}

Young stellar objects (YSOs) host a range of high-energy phenomena, pointing toward magnetic activity (e.g., 
\citealt{fem99,ben10}). Of the variability of YSOs across the electromagnetic spectrum, radio and X-ray variability are most 
closely related to magnetic activity, and indeed YSOs show strong, very short lived flares in 
these wavelength ranges. X-ray flares of this nature have been comparatively well-studied (e.g., \citealt{get08a,get08b}) 
whereas observations of greater numbers 
of non-thermal radio flares on timescales of hours or less have only 
recently become possible with sensitivity improvements, and thus 
comparably few such sources are known at this time. So far, the majority of observations of radio variability have been made on 
timescales of $\sim$days to years.

The first example of a YSO radio flare was reported by 
\citet{bower2003} toward the 
weak-line T Tauri star (class III YSO) GMR-A. The flux density at an 
observing frequency of 86~GHz rose by more than a factor of 5 within a few 
hours. X-ray monitoring of the source also revealed contemporaneous activity. Follow-up millimetre observations were performed 
by \citet{furuya2003}, who observed subsequent flare activity for $\sim$two weeks.
Another strong flare at an observing 
frequency of \mbox{90 GHz} was reported by \citet{massi2006} toward the weak-line T Tauri star 
V773~Tau, which also happened to be the first case with evidence for inter-
binary magnetic interaction.  \citealt{salter2008} observed a 3mm flare associated with the DQ Tau
 and showed evidence that these events may be expected for similar binary systems. Even the earliest evolutionary stages of a forming 
 star show activity: 
\citet{for08} reported a strong flare at 22~GHz 
toward a very deeply embedded protostar in the Orion BN/KL region. In all these cases, the radio emission is interpreted
 as non-thermal radiation.   
 
Giant flares from young stars are important both for understanding magnetic 
reconnection as well as for evaluating the impact of high energy photons on protoplanetary disks 
and atmospheric escape \citep[e.g.][]{guedel2014}.  The X-ray emission from 
flares, produced by hotter gas than quiescent X-ray emission, emits energetic 
photons that penetrate deep into the disk \citep{glassgold2004}, thereby 
affecting midplane ionization and chemistry \citep[e.g.][]
{rab2017,fraschetti2018}.  Variability has been detected in 
H$^{13}$CO$^+$ emission and attributed to flares \citep{cleeves2017}.  Flash-heating 
from X-rays is a possible explanation for producing chondrites 
\citep{shu1997} and abundances of calcium-rich inclusions seen in meteorites 
\citep{sossi2017}.  Once planets form, the ultraviolet and X-ray emission and 
cosmic rays produced by magnetic reconnection affects the atmospheric 
chemistry and enhances their escape \citep[e.g.][]{lammer2018}.

The correlation between X-ray and radio variability in YSOs has remained 
unclear, however, even though there is undoubtedly an underlying connection (e.g., 
\citealt{gue93}). Recently, \citet{for16,for17} analysed the 
incidence of order-of-magnitude flaring variability on short timescales in the 
Orion Nebula Cluster, using simultaneous radio (4.7 and 7.3~GHz) and X-ray observations. They found 
that there is a close connection only for a subset of this 
extreme radio and X-ray variability. 

In this paper, we present the detection of a submillimeter-wavelength outburst 
from the YSO \mbox{JW 566} in Orion, discovered in our the JCMT-Transient sub-mm 
monitoring program of nearby star-forming regions \citep{herczeg2017}.  This flare falls into the same category as the 
short-timescale, likely non-thermal radio flares described above.  
The short timescale for the decay of the flare rules out reprocessed accretion luminosity through thermal dust 
emission in the envelope \citep{johnstone2013} as the source of variability, which until this paper had been 
the type of variability uncovered within our survey \citep{yoo2017,mairs2017GBSTrans,johnstone2018},
even though this observation by itself does not necessarily imply entirely distinct emission mechanisms. 
This is the first such flare discovered 
at submillimeter wavelengths. In Section \ref{sec:obs}, we present 
details of our observations and data reduction methods. In Section 
\ref{sec:methods}, we introduce the methods used to detect the flare. In 
Section \ref{sec:source}, we display the \shortwave$\,$ and \longwave$\,$ 
images and construct a short timescale light curve of the flare. In Section 
\ref{sec:discuss} we discuss the results and compare the data presented in 
this paper to previous observations of \mbox{JW 566} at other wavelengths. 
Finally, in Section \ref{sec:summary}, we summarise our findings.

\section{Observations and Data Reduction} 
\label{sec:obs}

\subsection{SCUBA-2 Observations}

The JCMT Transient Survey (project code: M16AL001; 
\citealt{herczeg2017}) has been monitoring 
sub-mm emission from the Orion Molecular Cloud (OMC) 2/3 region, centered at 
(R.A., Decl.) = (05:35:33,$-5$:00:32, J2000), and seven other nearby star-
forming regions with a roughly monthly cadence since December 2015.
Each region is observed with SCUBA-2 \citep{holland2013} at \shortwave$\,$ and 
\longwave$\,$ 
simultaneously, with beam sizes of $9.8\arcsec$ and 
$14.6\arcsec$, respectively \citep{dempsey2013}. The images are constructed using $2\arcsec$ pixels at \shortwave$\,$ and $3\arcsec$ pixels at \longwave. 
The SCUBA-2 observing mode 
\textit{Pong1800} \citep{kackley2010} yields maps with a smooth 
sensitivity over a circular region with $30^\prime$ diameter.
Our integration times are set to ensure a consistent background noise level of 
$\sim10 \mathrm{\:mJy\:beam}^{-1}$ at \longwave$\,$ from epoch to epoch 
\citep{mairs2017}. At \shortwave, the submillimetre emission from the 
atmosphere has a much more significant effect on the the data, producing 
a range of noise levels that span more than an order of magnitude. 

Table \ref{tab:obssum} provides a log for our observations of 
OMC 2/3.  The flare is detected in our observations of 2016-11-26, beginning 
at MJD=57718.453 and ending at MJD=57718.474. We also include in our 
analysis of \mbox{JW 566} engineering/commissioning observations (project 
code M16BEC30) obtained on 2016-11-20 (UT), 6 days before the flare, which 
were obtained to assess the health of SCUBA-2 after maintenance.  
This engineering observation is used only to determine the 
detectability of \mbox{JW 566}, and is excluded from our co-added images and 
global analysis.

\begin{deluxetable*}{cccccc}[b!]
\tablecaption{A summary of the OMC 2/3 observations\tablenotemark{a} performed by the JCMT Transient Survey to date. Bold text highlights the date of the flare event (see Section \ref{sec:discuss}).}
\tablecolumns{6}
\tablenum{1}
\tablewidth{0pt}
\tablehead{
\colhead{UT Date} &
\colhead{Scan\tablenotemark{b}} &
\colhead{$\tau_{\mathrm{225}}$\tablenotemark{c}} & 
\colhead{Airmass} &
\colhead{\longwave$\,\sigma_{rms}$\tablenotemark{d}} &
\colhead{\shortwave$\,\sigma_{rms}$\tablenotemark{d}}\\
\colhead{(YYYY-MM-DD)} & \colhead{} & \colhead{} & \colhead{} & \colhead{(mJy beam$^{-1}$)} & \colhead{(mJy beam$^{-1}$)}
}
\startdata
2015-12-26 & 36 & 0.11 & 1.13 & 9.9 & 411.2\\
2016-01-16 & 19 & 0.06 & 1.21 & 7.7 & 81.3\\
2016-02-06 & 12 & 0.04 & 1.36 & 9.7 & 73.4\\
2016-02-29 & 11 & 0.04 & 1.16 & 9.4 & 67.4\\
2016-03-25 & 15 & 0.06 & 1.27 & 8.7 & 101.8\\
2016-04-22 & 11 & 0.05 & 1.72 & 9.1 & 126.4\\
2016-08-26 & 20 & 0.11 & 1.26 & 12.1 & 398.6\\
2016-11-20\tablenotemark{e} & 20 & 0.084 & 1.31 & 7.9 & 174.8\\
\textbf{2016-11-26} & \textbf{52} & \textbf{0.06} & \textbf{1.11} & \textbf{7.7} & \textbf{85.2}\\
2017-02-06 & 21 & 0.12 & 1.15 & 9.8 & 392.5\\
2017-03-18 & 12 & 0.10 & 1.10 & 9.1 & 255.6\\
2017-04-21 & 22 & 0.09 & 1.41 & 10.4 & 382.8\\
2017-07-08 & 73 & 0.05 & 1.24 & 9.4 & 72.1\\
2017-08-12 & 56 & 0.07 & 1.60 & 8.5 & 187.1\\
2017-09-12 & 54 & 0.10 & 1.32 & 9.0 & 280.3\\
2017-10-21 & 47 & 0.09 & 1.11 & 8.0 & 164.8\\
2017-11-18 & 46 & 0.06 & 1.10 & 9.5 & 77.6\\
2017-12-24 & 46 & 0.07 & 1.36 & 8.5 & 148.5\\
2018-02-08 & 18 & 0.09 & 1.20 & 8.0 & 184.6\\
2018-03-08 & 15 & 0.11 & 1.64 & 10.4 & 780.1\\
\enddata
\tablenotetext{a}{The target identifier in the JCMT archive is OMC2-3.}
\tablenotetext{b}{The observation number.}
\tablenotetext{c}{$\tau_{\mathrm{225}}$ is the zenith opacity of the atmosphere at 225 GHz.}
\tablenotetext{d}{Noise measurements are based on the pixel variances in a central 900$\arcsec$ radius of the Gaussian smoothed map.}
\tablenotetext{e}{This data was taken as part of an engineering and commissioning project (project ID: M16BEC30) and not part of the JCMT Transient Survey.}
\label{tab:obssum}
\end{deluxetable*}

\subsection{Data Reduction and Flux Uncertainties}
\label{subsec:DR}

The data reduction was carried out using the iterative map-making 
software, {\sc{makemap}} \citep[see ][ for details]{chapin2013}, which is part 
of {\sc{starlink}}'s \citep{currie2014} 
Submillimetre User Reduction Facility ({\sc{smurf}}) package 
\citep{jenness2013}. The specific data reduction and image 
calibration techniques used by the JCMT Transient Survey are 
described in detail by \cite{mairs2017} (reduction \textit{R3} with no 
relative flux calibration applied\footnote{To
suppress the false detections of noise spikes or artificial structure, the images are smoothed 
with Gaussian kernels with FWHM values of
$4\arcsec$ and $6\arcsec$ at \shortwave$\,$ and \longwave, respectively (two pixels in each case).}). 
Briefly, to perform a \textit{Pong1800} observation, 
the telescope continually scans across the sky, observing each 
location at a variety of position angles. This technique allows 
for the modeling and subtraction of the large-scale, bright, 
variable atmosphere at submillimetre wavelengths. The continual scanning across the sky also provides a time-series, which is exploited in this paper to measure light curves during our observation. 

To optimize the extraction of compact sources, we applied a stringent
spatial filter during the data reduction to suppress signal on scales 
$>200\arcsec$.  
A mask was used to define significant astronomical 
emission in the image which provided additional constraints to {\sc{makemap}} 
and aided in the background subtraction 
\citep[see ][]{chapin2013,mairs2015}.  
The OMC 2/3 region is particularly difficult to mask due to the large amount 
of extended emission, though all persistent point sources are well accounted 
for in the regular analysis pipeline.  The source with a strong flare, 
\mbox{JW 566}, resides in a region of \textit{negative bowling}, a section 
of the image with artificially low values just outside the boundaries of 
bright, extended emission, due to the application of the stringent spatial 
filter. The external mask was therefore adjusted to include the bright event 
associated with \mbox{JW 566} and the data reduction was re-run for all 
epochs to ensure the flux of the source was well recovered. 

In addition, all Transient Survey epochs, excluding 2016-11-26 (the date of 
the flare event), were then co-added and subtracted from the 2016-11-26 data 
to produce residual flux maps at \shortwave$\,$ and \longwave. The fluxes of \mbox{JW 566} quoted in Section 
\ref{sec:source} were measured in the residual maps to suppress  
any background structure present in the image. At \longwave, the 
background level of the co-add at the location of \mbox{JW 566} is 
$2\mathrm{\:mJy\:beam}^{-1}$, which is insignificant relative to the peak 
flux measurement of the source. We adopt the JCMT standard flux uncertainty 
value of $10\%$ for \longwave$\,$ observations (\citealt{dempsey2013}; Mairs et al, in prep.). 
At \shortwave, however, \mbox{JW 566} is located within a negative 
bowl in the co-added image, with a background level of 
$-100\mathrm{\:mJy\:beam}^{-1}$. 
The standard deviation of the mean negative bowl depth within the region of \mbox{JW 566} for observations with similar background noise (450 $\mu\mathrm{m}\: \sigma_{rms} < 120\mathrm{\:mJy\:beam}^{-1}$) is $77\mathrm{\:mJy\:beam}^{-1}$ (see Table \ref{tab:obssum}). 
We therefore combine the JCMT standard flux uncertainty 
value at \shortwave$\,$ of $15\%$ (\citealt{dempsey2013}; Mairs et al, in prep.) with the uncertainty in the  
negative bowl depth for an 
uncertainty of 21\% for \mbox{JW 566}.

The fluxes described in this paper ignore any emission from $^{12}$CO J=3--2, 
which is located within the 850 $\mu$m filter of SCUBA-2  \citep[for a 
discussion on the effects of CO contamination, see][]
{drabek2012,coude2016,parsons2018}. In the JCMT Gould Belt Survey image of 
$^{12}$CO emission obtained with the Heterodyne Array Receiver Programme 
(HARP) \citep{buckle2009}, weak CO emission is dispersed across the location 
of \mbox{JW 566} with no significant compact structure, and has a negligible 
effect on the continuum emission described in this paper.

\subsection{Subdividing the Raw Data}

Raw SCUBA-2 data are comprised of the power received at the 
focal plane over time, with separate integrations read out and saved in 
$\sim36\mathrm{\:sec}$ intervals, some of which include observations of \mbox{JW 566}. 
In Section \ref{lightcurve}, we subdivide this data stream into nine shorter 
integrations based on when the telescope passed over \mbox{JW566}.
These subdivisions are reconstructed into individual 
images using the same external mask and data reduction parameters 
as for the full integration.

\section{Searching for Variability of Faint Sources}
\label{sec:methods}

At the time of writing, the JCMT Transient Survey has obtained 
nearly 3 years of data across 8 star-forming regions.
\cite{johnstone2018} analyzed source variability in all \longwave$\,$ images 
taken throughout the first 18 months of the survey (December 2015 
through May 2017).  The 1643 sources in that analysis were selected by 
identifying compact emission peaks with a brightness 5 times higher than the 
noise in the co-added maps of each region in our survey.  Since that time, 
these same sources have been tracked with an automated 
pipeline that measures the flux soon after the data are obtained and compares 
it with past observations.
The automated version of the \cite{johnstone2018} analysis, however, relies on 
the initial detection 
of a source in the co-added image of a given target field. 
Therefore, the pipeline is only sensitive to short timescale 
burst events that are strong enough to be detected at a 
level of $5\sigma_{rms}$ in the co-add.  We are now in the process of further 
evaluating transient variability by searching for any sources that might have 
appeared in only a single image (Lalchand et al.~in prep). 
Following the methods of \cite{johnstone2018}, we use the 
{\sc{JSA\_CATALOGUE}} program (found in {\sc{starlink}}'s 
{\sc{picard}} package \citealt{sun265}) to optimise and 
run the {\sc{fellwalker}} \citep{berry2015} source detection 
algorithm to identify compact, peaked, continuum emission 
structures. While \cite{johnstone2018} used this strategy to 
produce a source catalogue from the co-added image of a field, 
we repeat their procedure for each individual epoch in the \mbox{OMC 
2/3} region. In this way, we compare the catalogues 
generated for each epoch with the catalogue generated for the 
co-add and identify sources that appeared in individual 
observations, but not in the averaged image. We refer to these 
sources as \textit{candidate transients}.

\begin{figure}
\centering
\includegraphics[width=1.0\textwidth]{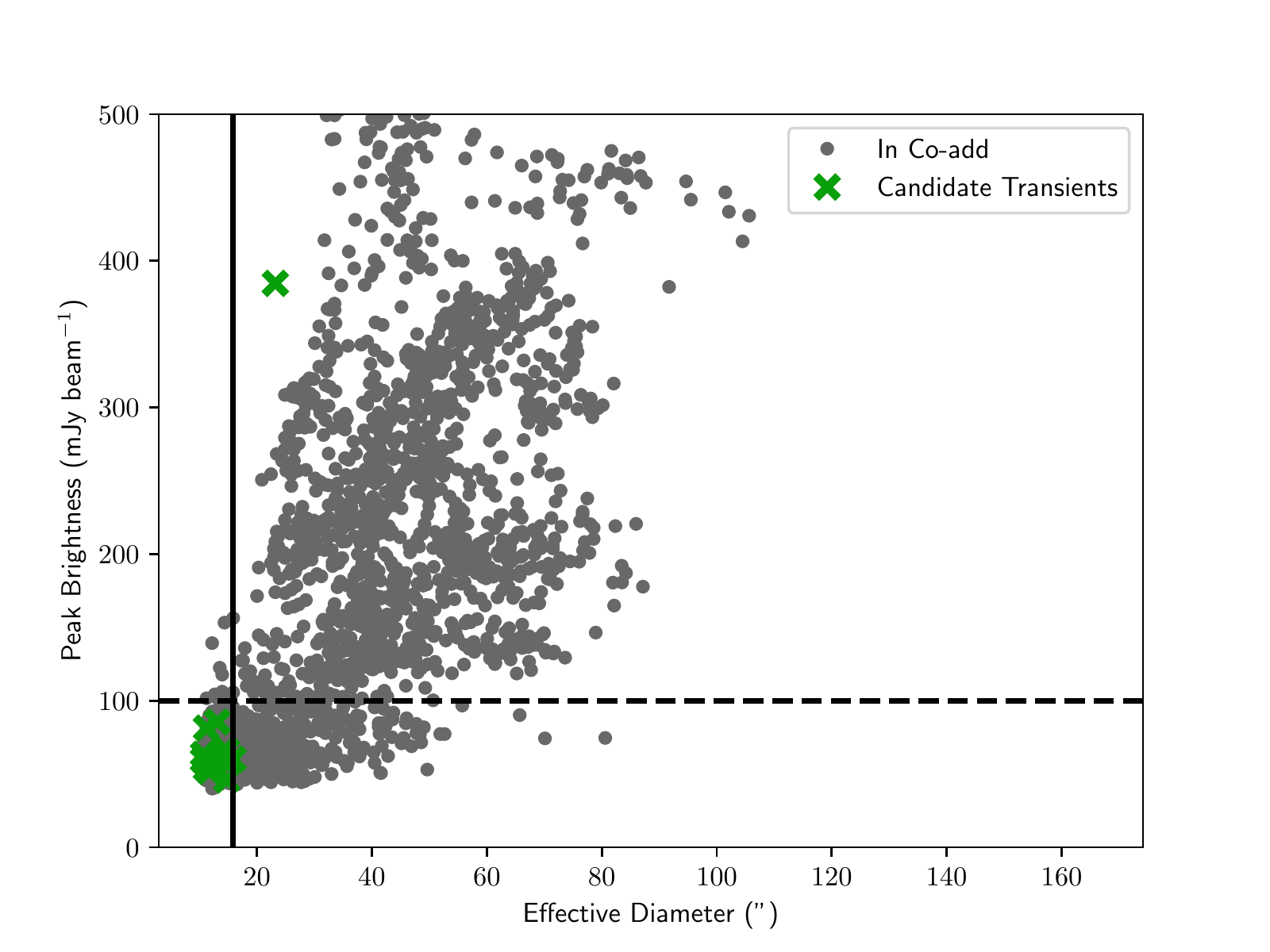}
\caption{The \longwave$\,$ peak brightness versus the effective diameter (assuming a circular projection) of all sources identified in all 19 OMC 2/3 epochs. Grey circles indicate sources identified in the co-add that are being analysed by the current pipeline. Green X's indicate candidate transient sources which do not appear in the co-add (at the level of 5$\sigma_{rms}$) but do appear in at least one epoch. A horizontal (dashed) line has been drawn at $100 \mathrm{\:mJy/beam} \sim 10\sigma_{\mathrm{rms}}$. The vertical (solid) line represents the effective \mbox{850 $\mu$m} beam FWHM after Gaussian smoothing.}
\label{fig:VarDetection}
\end{figure}

Figure \ref{fig:VarDetection} shows the \longwave$\,$ source peak  
(the maximum pixel value in each source footprint) of all the 
sources detected in the OMC 2/3 co-add (grey) versus the effective 
diameter.  
The effective diameter of a source is calculated by measuring the total area
in the identified clump footprint and assuming a projected circular symmetry. Since this method takes into account the full clump size (and not the 
brightness profile), unresolved sources will appear larger than the beam FWHM while false detections due to residual correlated noise in the final 
image will appear smaller than the beam FWHM. 
The circular symmetry assumption holds well for compact 
(approximately beam-sized) objects.
Also included (green) are the sources detected only in individual epochs. 
Several faint, spurious sources that do not appear at the level 
of 5$\sigma_{rms}$ in the co-add and are smaller than the beam
are detected in single epochs.  A full, multi-region analysis of candidate transients like these will 
be presented by Lalchand et al.~(in prep).   

One unresolved\footnote{A Gaussian fit at the position of the significant candidate transient source in the smoothed, \longwave$\:$ image results in a Full Width 
at Half Maximum value of 15.5$\arcsec$ averaged over the vertical and horizontal directions. The effective beam size after smoothing is 15.8$
\arcsec$.} source, however, is a clear 
outlier. This source is detected at (R.A., Decl.) = 
(5:35:17.94,$-5$:16:11) on 2016-11-26 (UT), with an initial \longwave$\,$ source peak of 
$384\mathrm{\:mJy\:beam}^{-1}$ in an observation 
that had a background noise level of $\sigma_{\mathrm{rms}} 
= 9.74\mathrm{\:mJy\:beam}^{-1}$ (SNR = 39). This flux, however, is underestimated as it was 
measured before the image was re-reduced with an appropriate mask (see Section 
\ref{subsec:DR}). The 2016-11-26 epoch is the only image with a detection of this source.  The source was not identified in our initial
measurements of variability from the first 11 epochs of the survey \citep{johnstone2018} because negative bowling in the region of \mbox{JW 566} 
led to an average source brightness of $\sim0.8\mathrm{\:mJy\:beam}^{-1}$ at the peak pixel location.

The position of this candidate transient peak  
is within $2.4\arcsec$ of the 
position of \mbox{JW 566} \citep{jw566}, a 
K7+M1.5 T Tauri binary system with a projected separation of $0\farcs86$ \citep{daemgen2012}. It has been classified as a ``Disk'' by \cite{megeath2012} based on its mid-infrared colours. 

\section{A Sub-mm Flare of JW 566} 
\label{sec:source}

\subsection{Detecting the flare at \longwave}
\label{sec:850detection}

\begin{figure}
\centering
\includegraphics[width=0.75\textwidth]{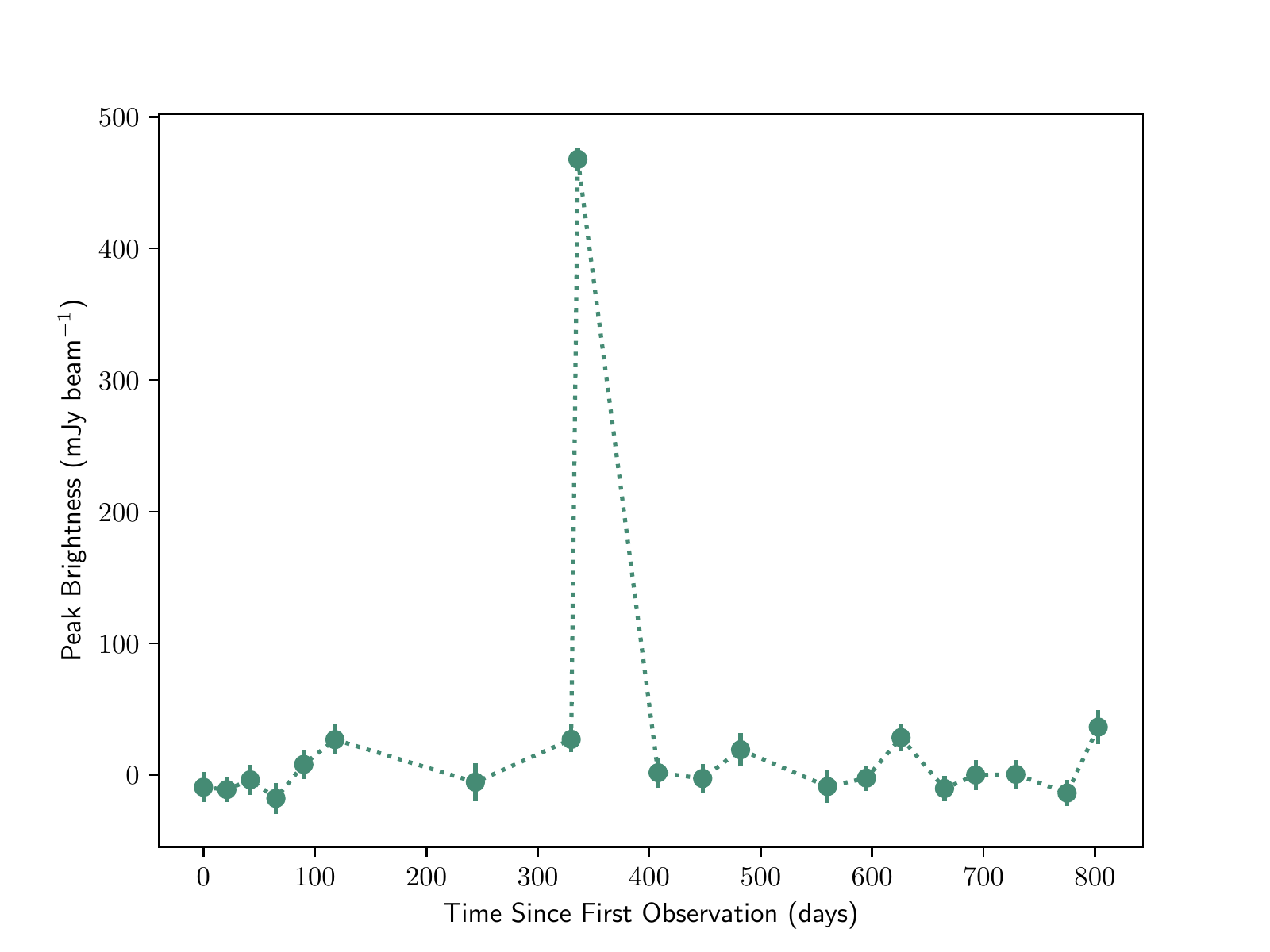}
\caption{The \longwave$\,$ light curve of \mbox{JW 566} over all observed epochs (see Table \ref{tab:obssum}).}
\label{fig:LightcurveAllEpochs}
\end{figure}

\begin{figure*}
\gridline{\fig{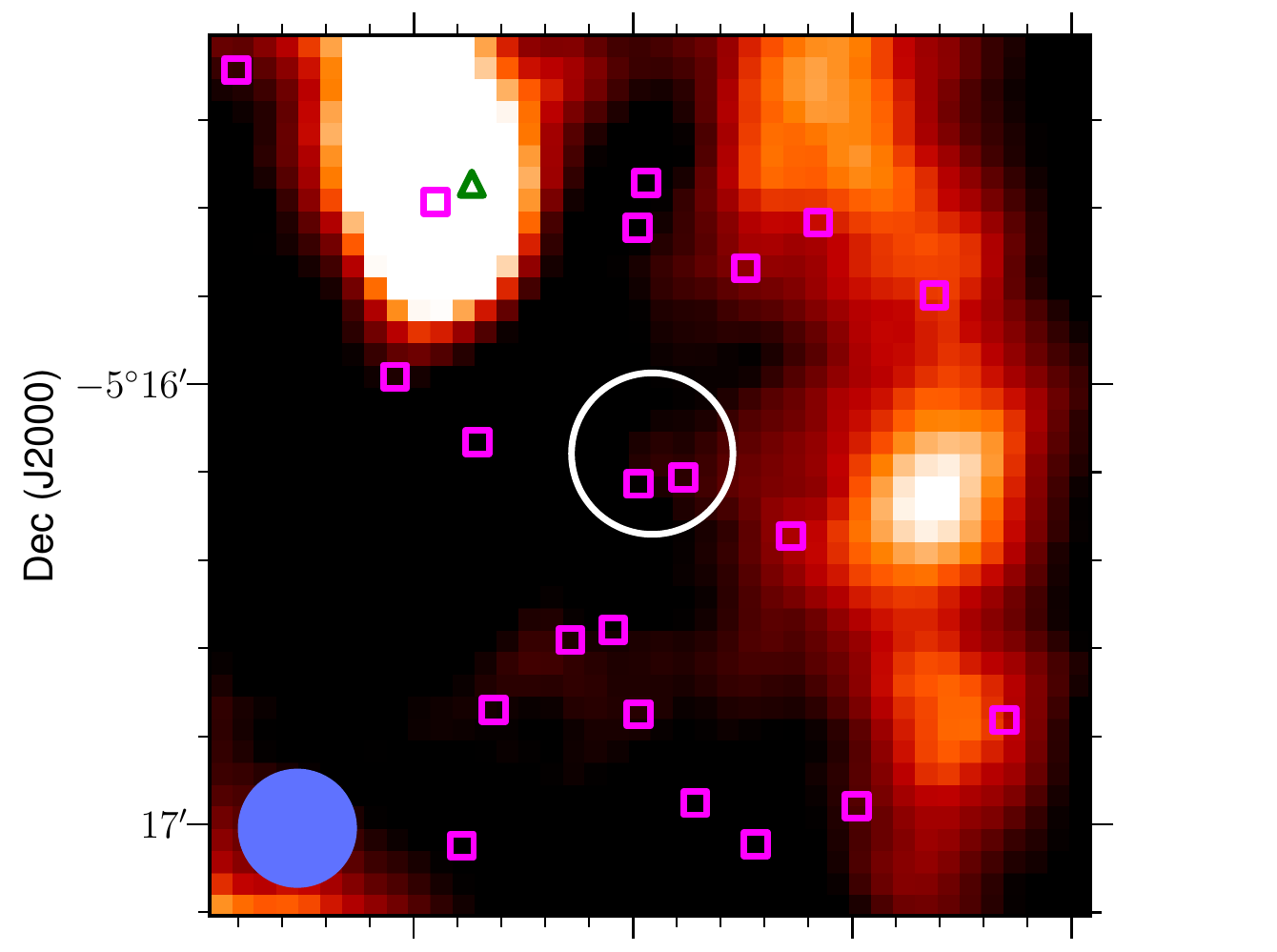}{0.48\textwidth}{(a) \longwave$\:$ 2016-11-20 (UT).}
          \fig{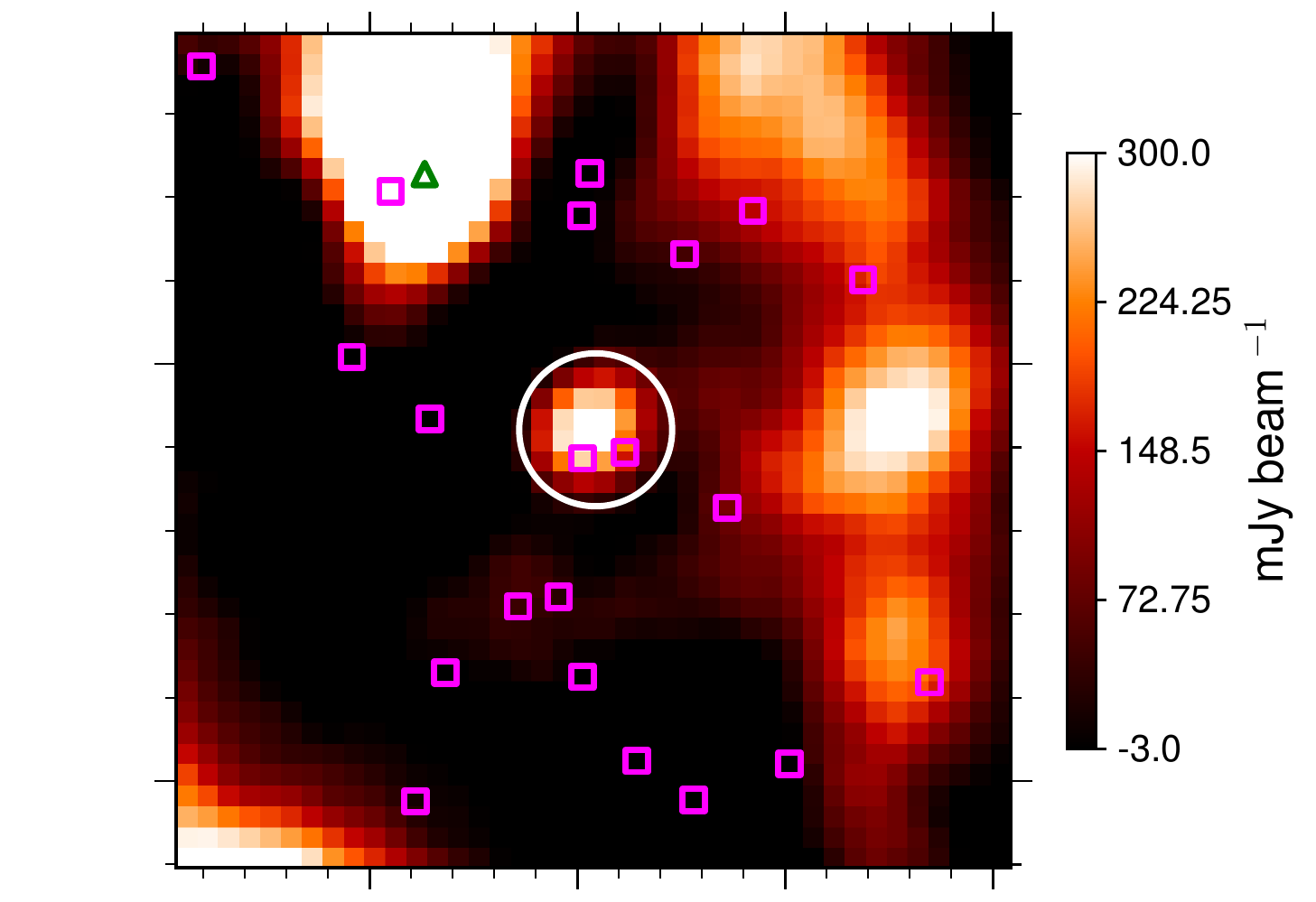}{0.525\textwidth}{(b) \longwave$\:$ 2016-11-26 (UT).}
          }
\gridline{\fig{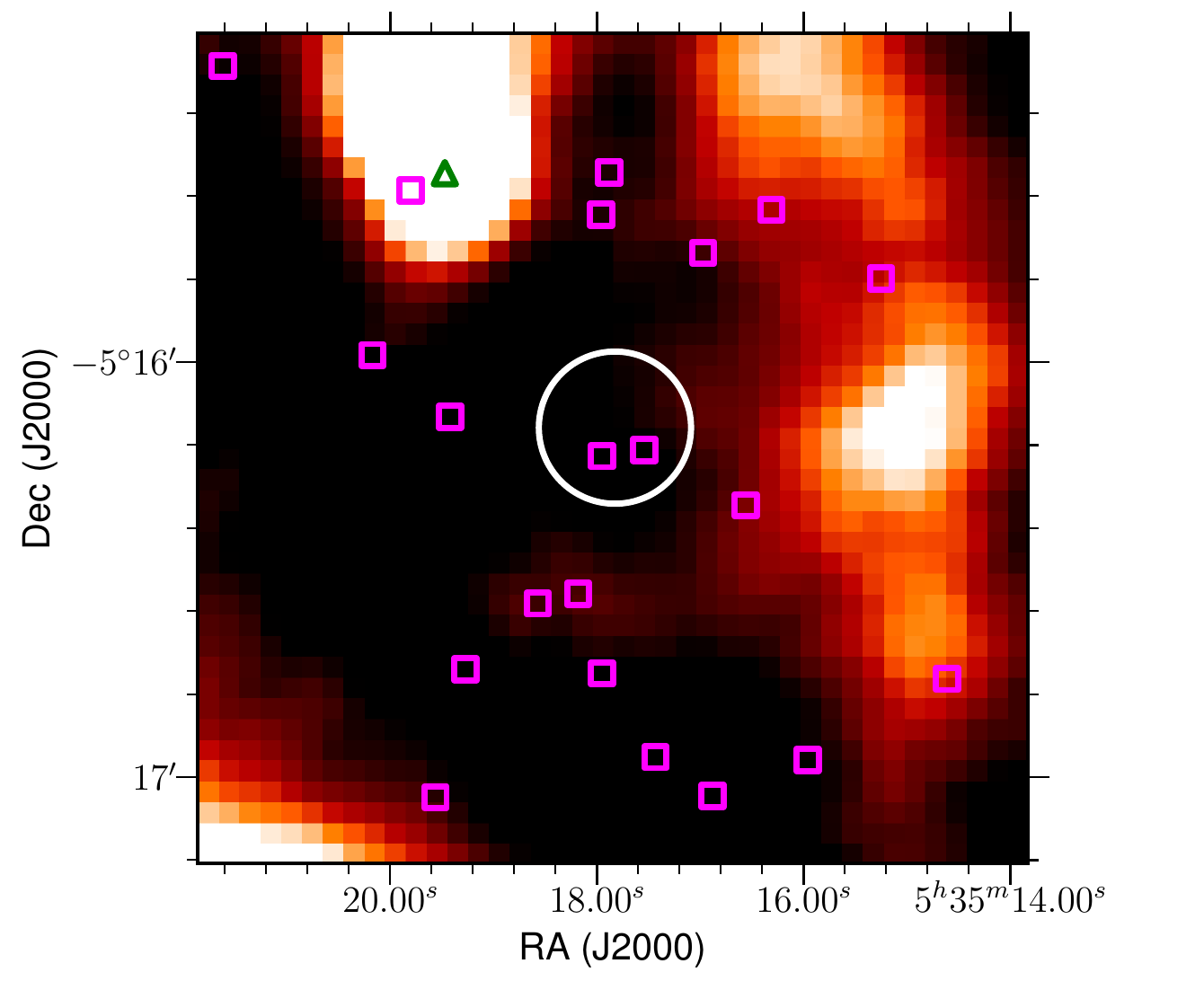}{0.48\textwidth}{(c) \longwave$\:$ 2017-02-06 (UT).}
          \fig{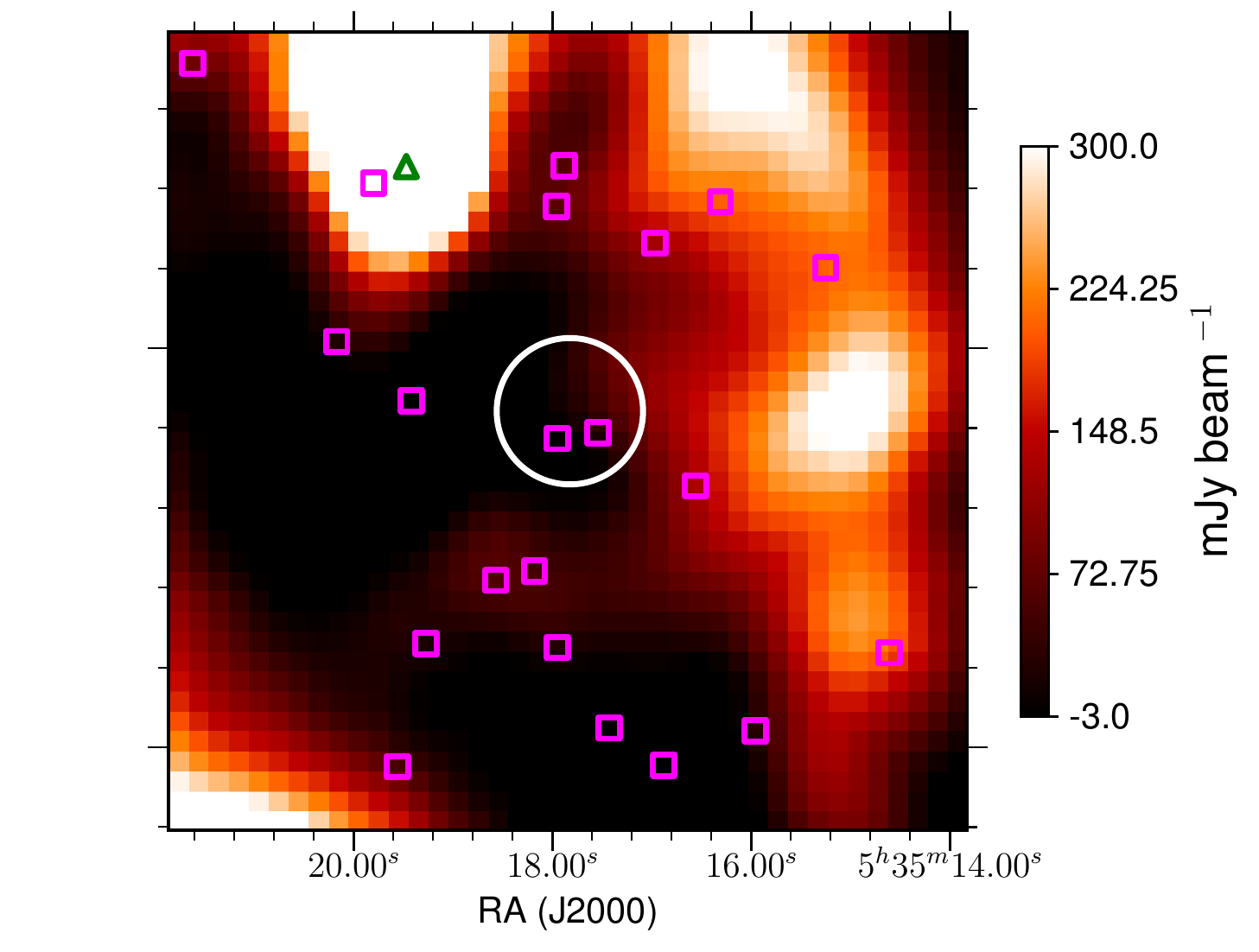}{0.525\textwidth}{(d) The co-add of all \longwave$\:$ epochs not including 2016-11-26.}
          }
\caption{\longwave$\,$ observations of JW 566. Image (a) was observed as part of an engineering program carried out by JCMT staff on the night of 2016-11-20 (UT); the beam is shown in blue. Images (b) and (c) are two consecutive epochs taken as part of the JCMT Transient Survey; the flare occurred on 2016-11-26 [image (b)]. The green triangle represents the position of a known protostar while the magenta squares mark the positions of known Class II YSOs \protect{\citep{megeath2012}}. The white circle shows the location of JW 566. Image (d) is a co-add of all \longwave$\,$ epochs not including 2016-11-26. }
\label{fig:newvar}
\end{figure*}

Out of 20 epochs of SCUBA-2 imaging, 
only one (2016-11-26) shows bright, 
unresolved \longwave$\,$ emission at the location of \mbox{JW 566} 
(see Table \ref{tab:obssum} and Figure 
\ref{fig:LightcurveAllEpochs}). Figure \ref{fig:LightcurveAllEpochs} presents a light curve derived from extracting the value of the pixel at the peak 
location of \mbox{JW 566}\footnote{The small fluctuations 
are due to a slight amount of faint, extended emission in this region which is better recovered 
in some epochs but it is not associated with \mbox{JW 566}.}. The uncertainties are calculated by measuring the average pixel 
variance\footnote{Each SCUBA-2 map has an associated 
``variance map'' that records the variance of the bolometer signals contributing to each pixel.}
 in a $20\times20$ pixel box centered on the source. Figure \ref{fig:newvar} shows that emission is not detected at this position  
in the previous image obtained six days earlier, or in
 the subsequent image obtained three months later.  After co-adding all 18 
Transient Survey epochs without a detection, the source is still not 
detected, with a noise of $\sigma_{\mathrm{rms}} = 3\mathrm{\:mJy
\:beam}^{-1}$ in the co-added map (indicating an upper limit on the flux of $\sim3\sigma_{\mathrm{rms}} = 9\mathrm{\:mJy
\:beam}^{-1}$).  The source is also not detected in the co-add of the SCUBA-2 images 
obtained by the JCMT Gould Belt Survey \citep{wardthompson2007} in 
2011, with a sensitivity of $\sim 4\mathrm{\:mJy\:beam}^{-1}$ (Data Release 3; \citealt{kirk2018arxiv}).

The bright peak associated with \mbox{JW 566} is detected in the map 
obtained during a 31-minute observation.  The excess emission is 
consistent with an unresolved object at the (non-smoothed) 14.6$\arcsec$ 
resolution of the JCMT. The source is best detected in a residual map of the 
co-add [Figure \ref{fig:newvar}(d)] subtracted from the flare epoch 
[Figure \ref{fig:newvar}(b)] (after the images were re-reduced with 
the new mask; see Section \ref{subsec:DR}]).
The average brightness of the source during our observation is \mbox{$466\pm47\mathrm{\:mJy
\:beam}^{-1}$} (SNR = 48), as measured by fitting a Gaussian profile to 
the source in the residual map.


\subsection{Detecting the flare at \shortwave}
\label{sec:450detection}

\begin{figure*}
\gridline{\fig{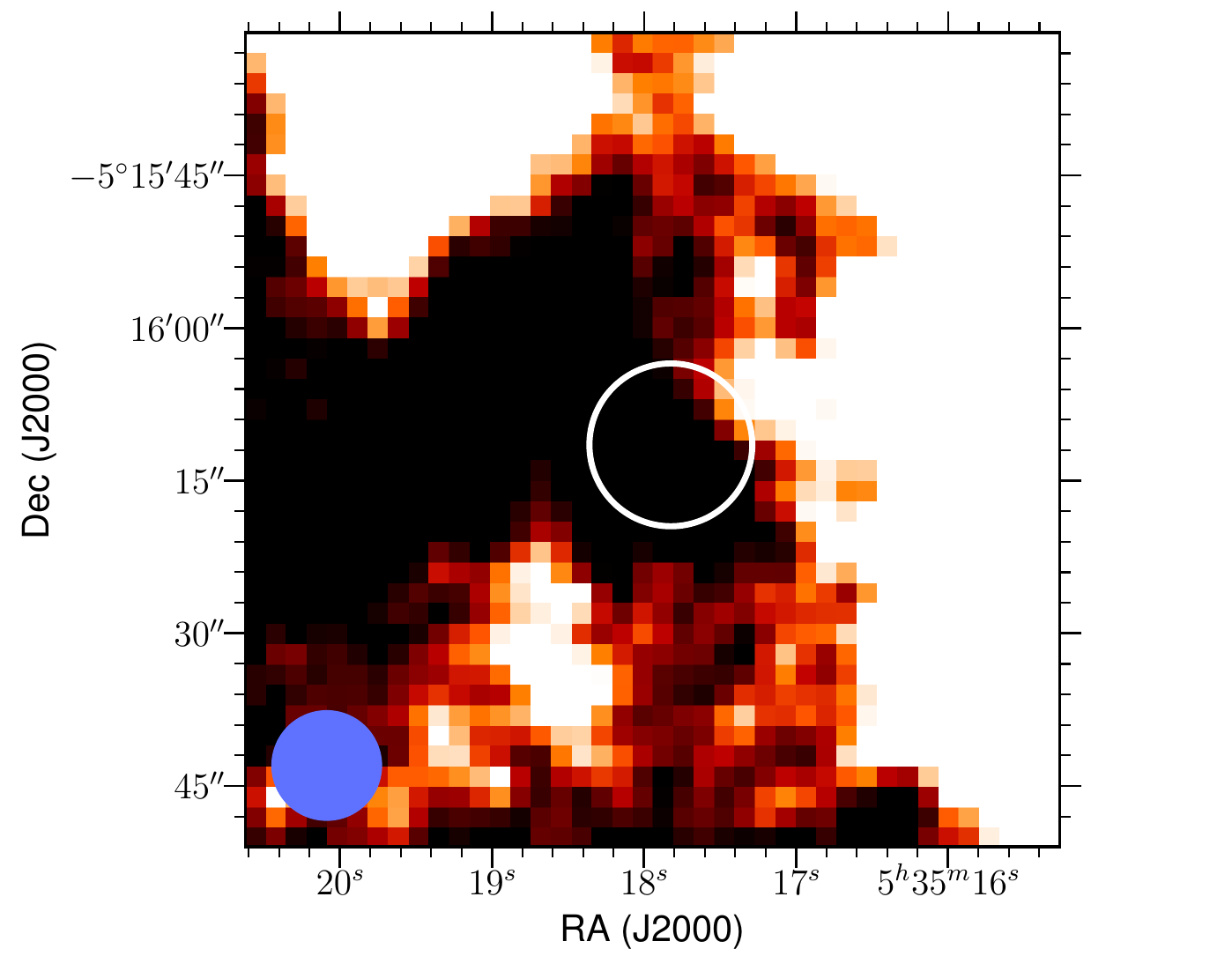}{0.32\textwidth}{(a) Co-added data (not including observation on 2016-11-26).}
          \fig{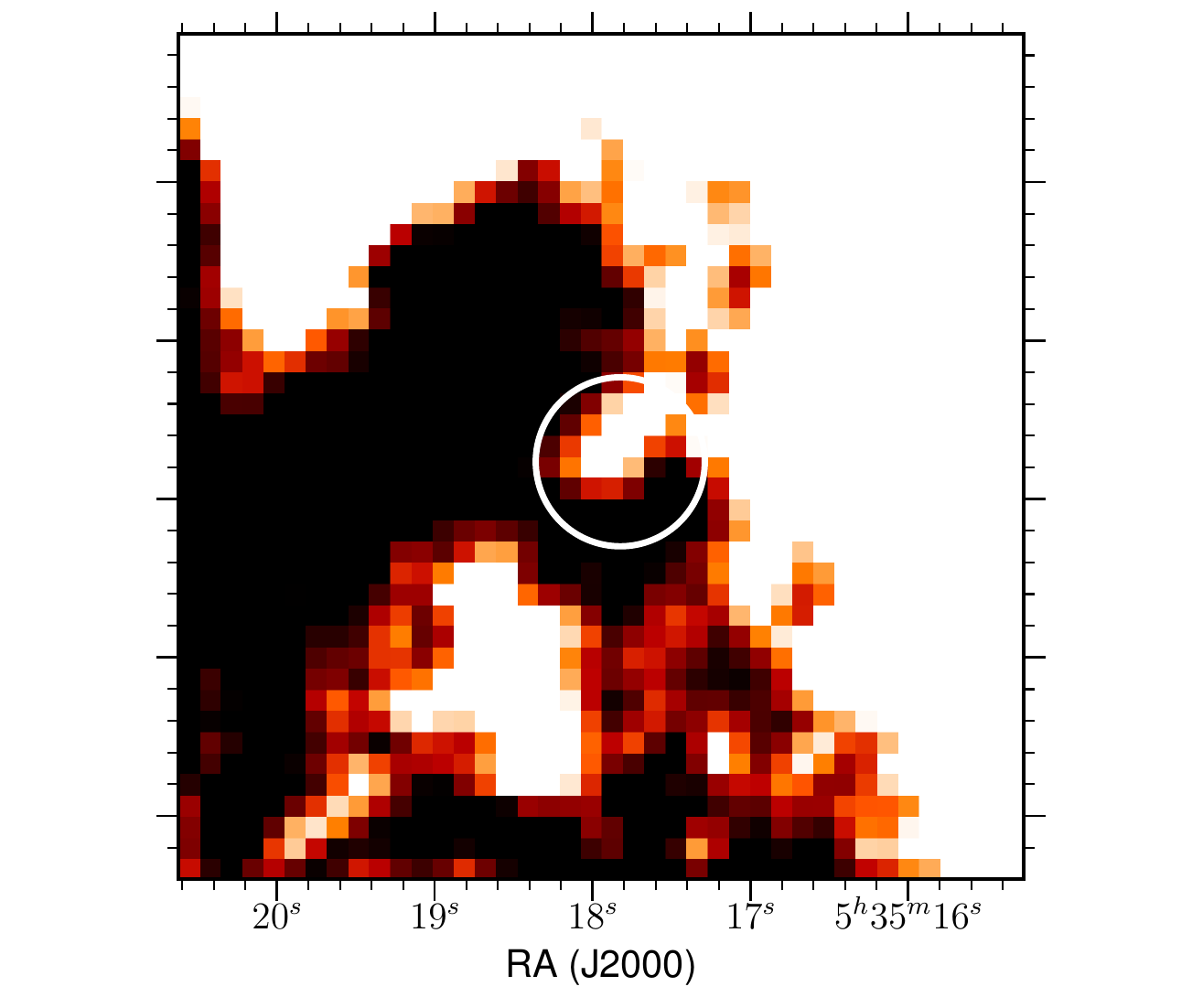}{0.305\textwidth}{(b) 2016-11-26 (UT).}
          \fig{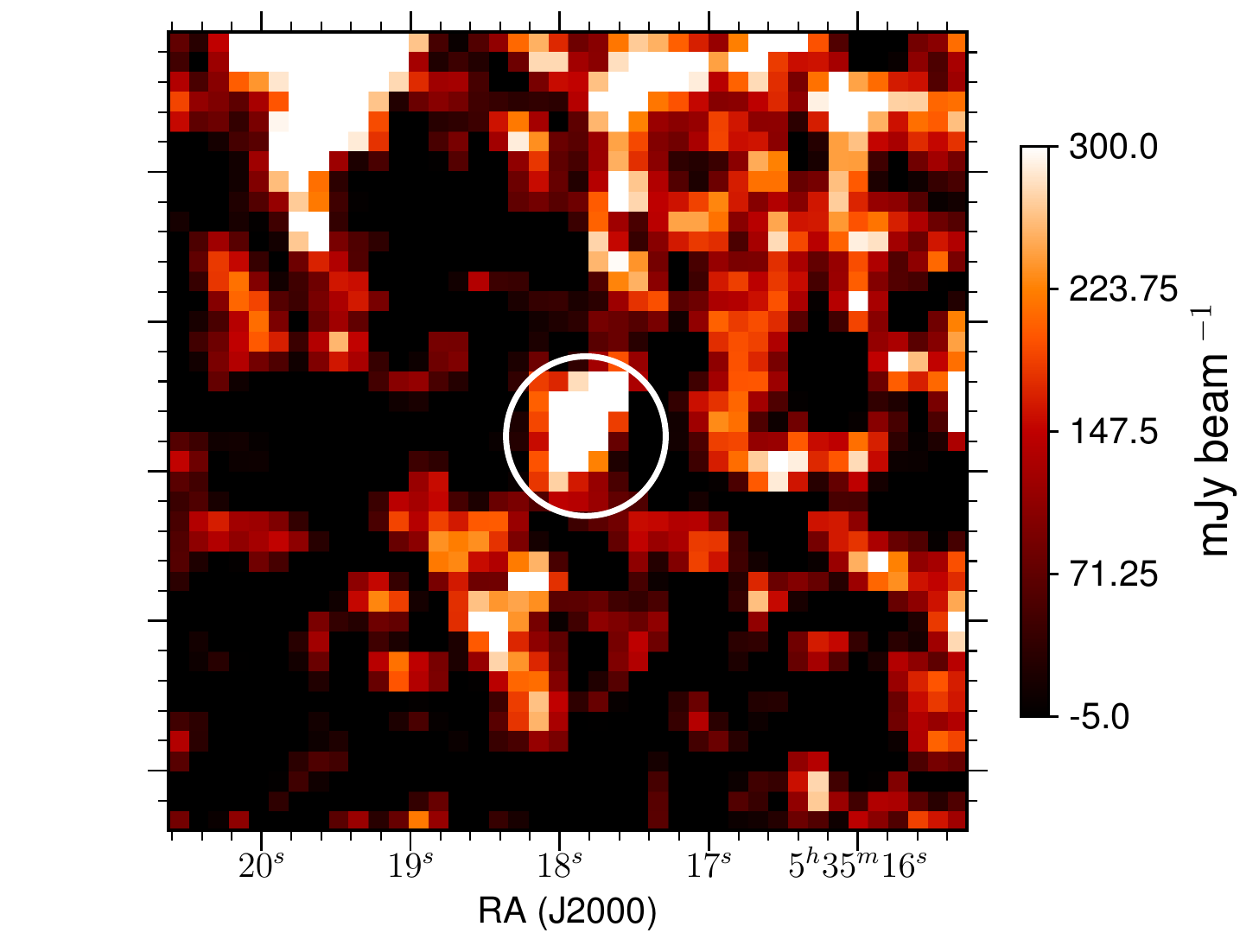}{0.335\textwidth}{(c) Observation on 2016-11-26 minus the co-add.} 
          }
\caption{\shortwave$\,$ observations of \mbox{JW 566}. The beam size is shown as a blue circle in image (a), the co-added data not including the 2016-11-26 epoch. Image (b) shows the 2016-11-26 epoch when the flare occurred. Image (c) shows the subtraction of image (a) from image (b). Compact structure is obvious in the residual.}
\label{fig:newvar450}
\end{figure*}

The brightness peak of \mbox{JW 566} is also detected in the simultaneous \mbox{450 $\mu$m} images obtained with SCUBA-2.  
The precipitable water vapor was low on the night of the flare, leading to a noise 
level of $\sigma_{\mathrm{rms}} = 85\mathrm{\:mJy\:beam}^{-1}$.  Figure \ref{fig:newvar450} presents the co-
added data in image (a), excluding the 2016-11-26 epoch, the 
2016-11-26 (flare) epoch in image (b), and a subtraction of image (a) from 
image (b) in image (c).  

As in the case of the 
\longwave$\,$ observations, there is no indication of 
significant emission correlated with the position of 
\mbox{JW 566} in any other SCUBA-2 image. A 2-dimensional Gaussian profile fit to the residual \shortwave$\,$ image yields a source peak of \mbox{$500\pm107\mathrm{\:mJy\:beam}^{-1}$} (the detection has a SNR=6; see Section \ref{subsec:DR} for more information about the uncertainty).

\subsection{Minute to Minute Variability at \longwave}
\label{lightcurve}

Since JW 566 is not detected in the engineering data, 6 days prior to 
the flare, the source must vary on timescales shorter than one week.  In this 
section, we analyze the light curve of the bright emission 
peak in nine separate intervals within the 31.12 minute integration 
of 2016-11-26.  In each interval, the source is detected with a SNR 
between 5 and 25  
and is fit with a 2-dimensional Gaussian profile to measure the 
source peak.  

\begin{figure*}
\gridline{\fig{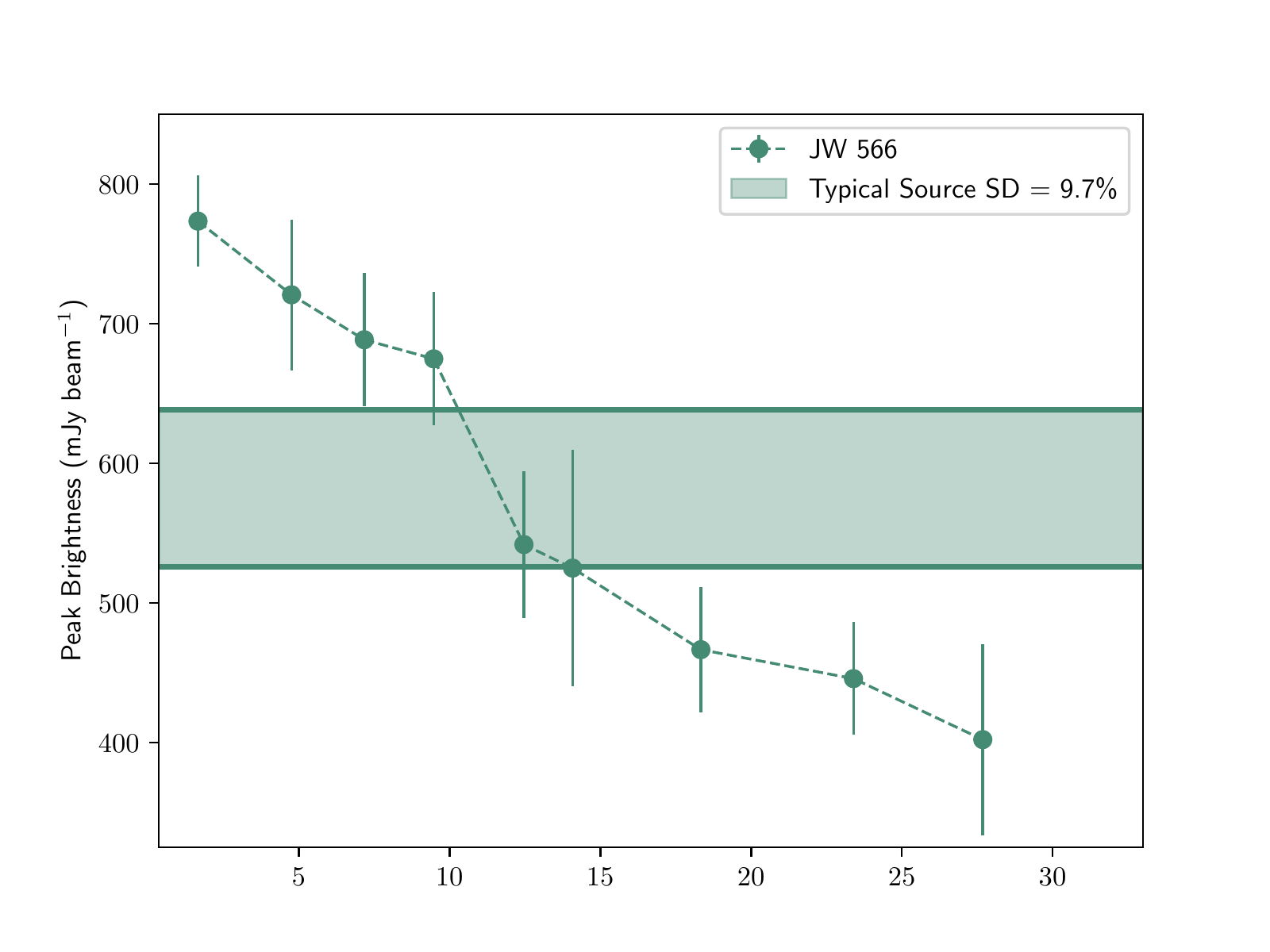}{0.427\textwidth}{}
          \fig{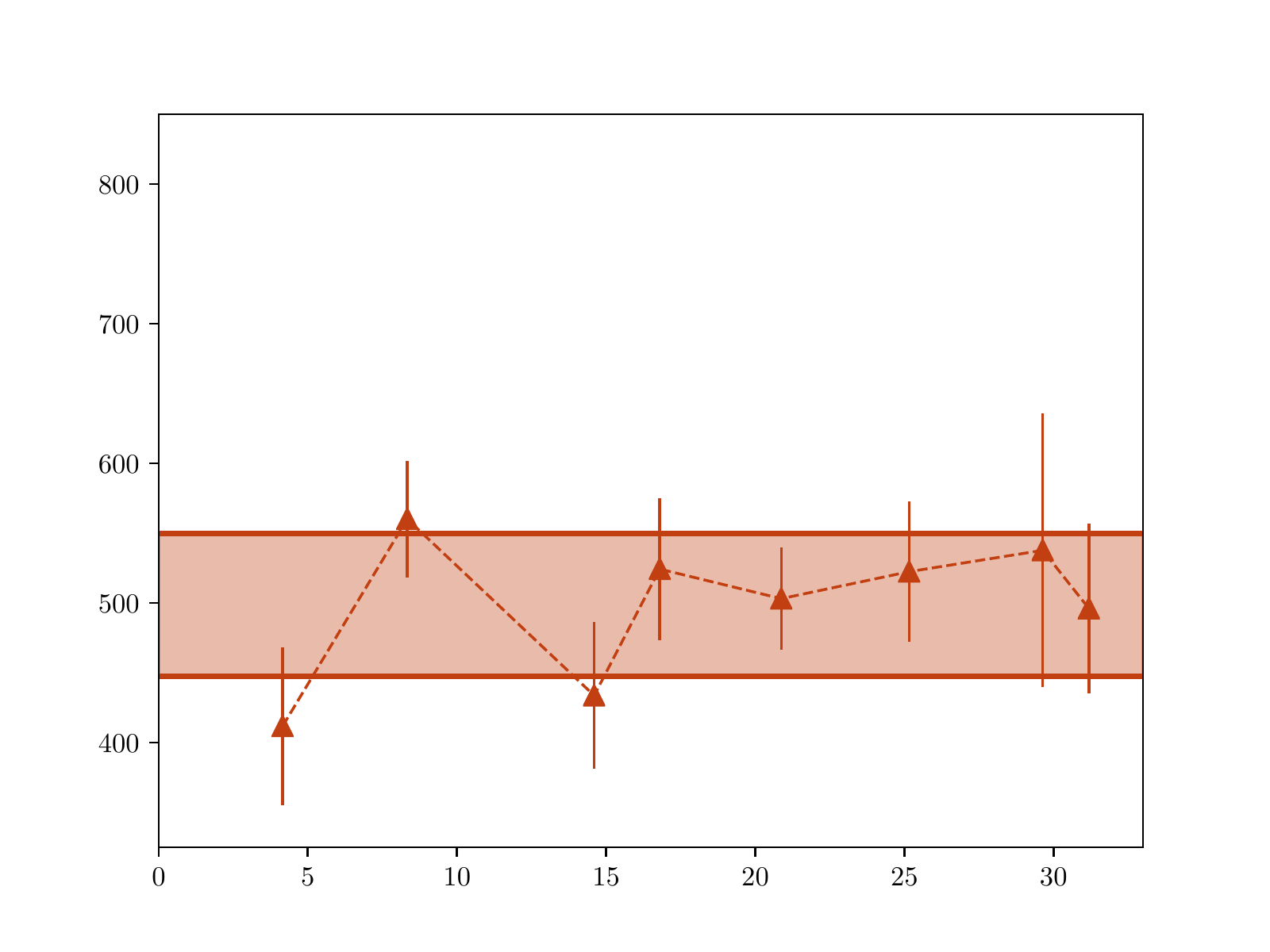}{0.427\textwidth}{}
          }
\gridline{\fig{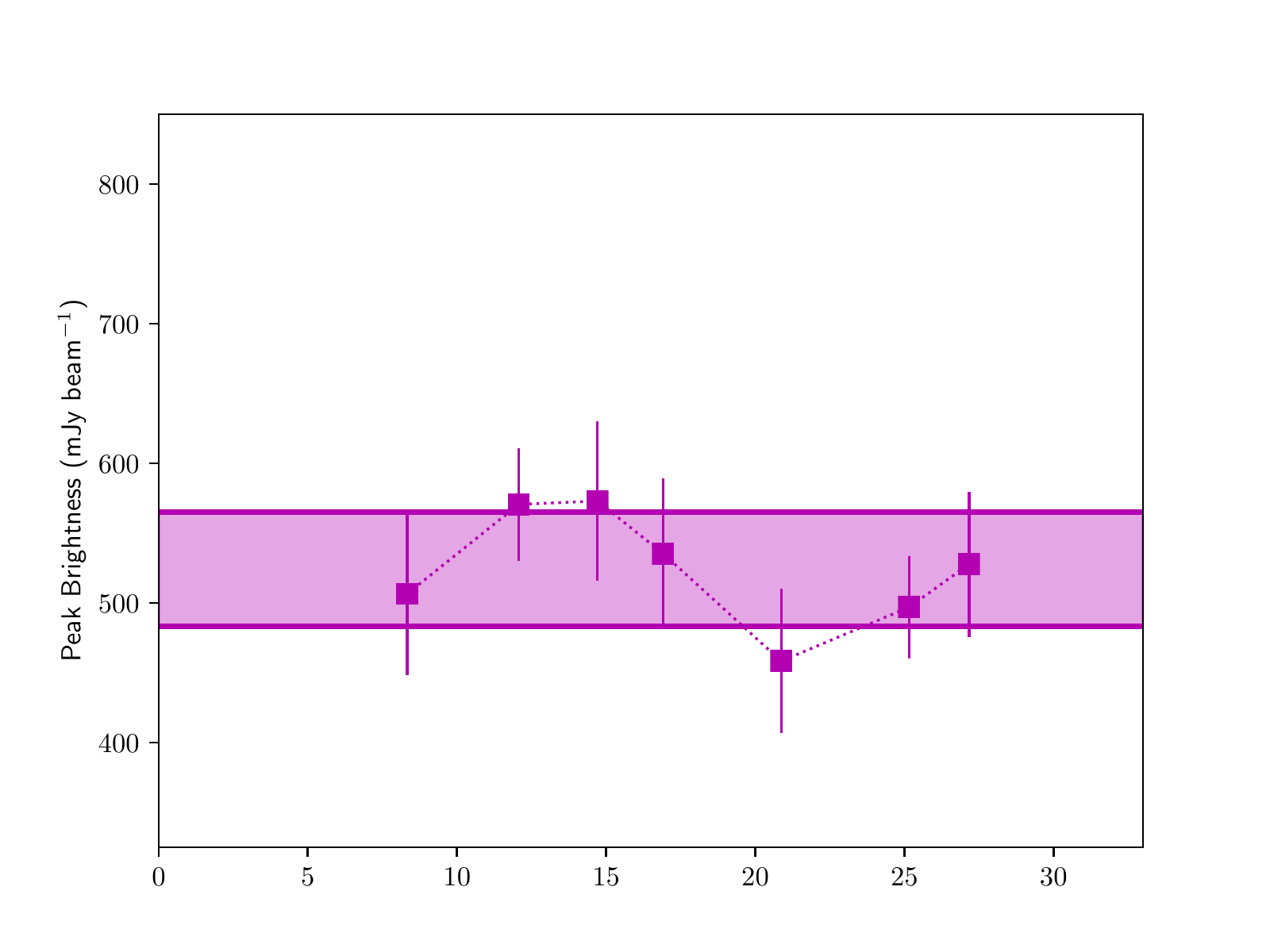}{0.427\textwidth}{}
          \fig{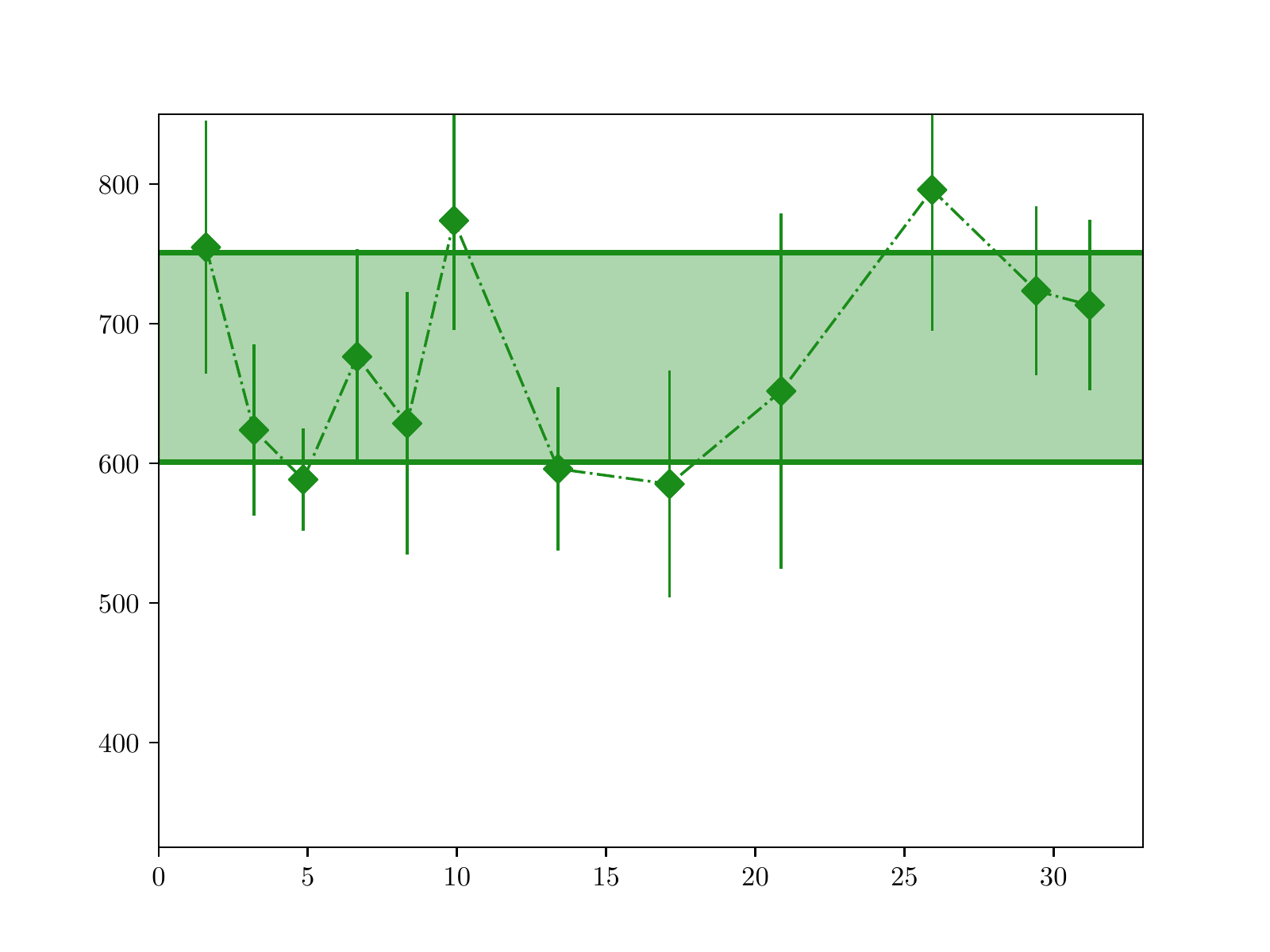}{0.427\textwidth}{}
          }
\gridline{\fig{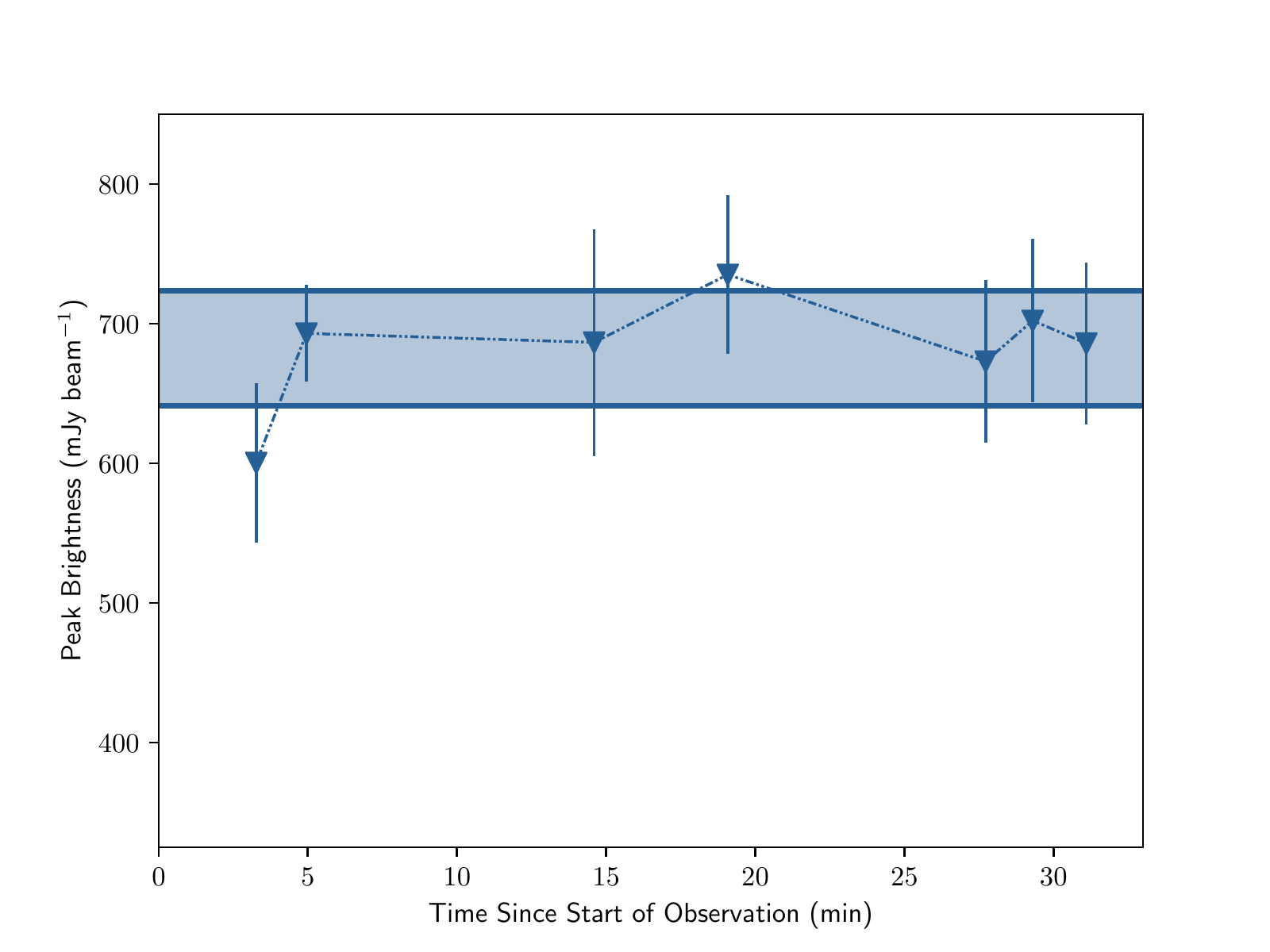}{0.427\textwidth}{}
          \fig{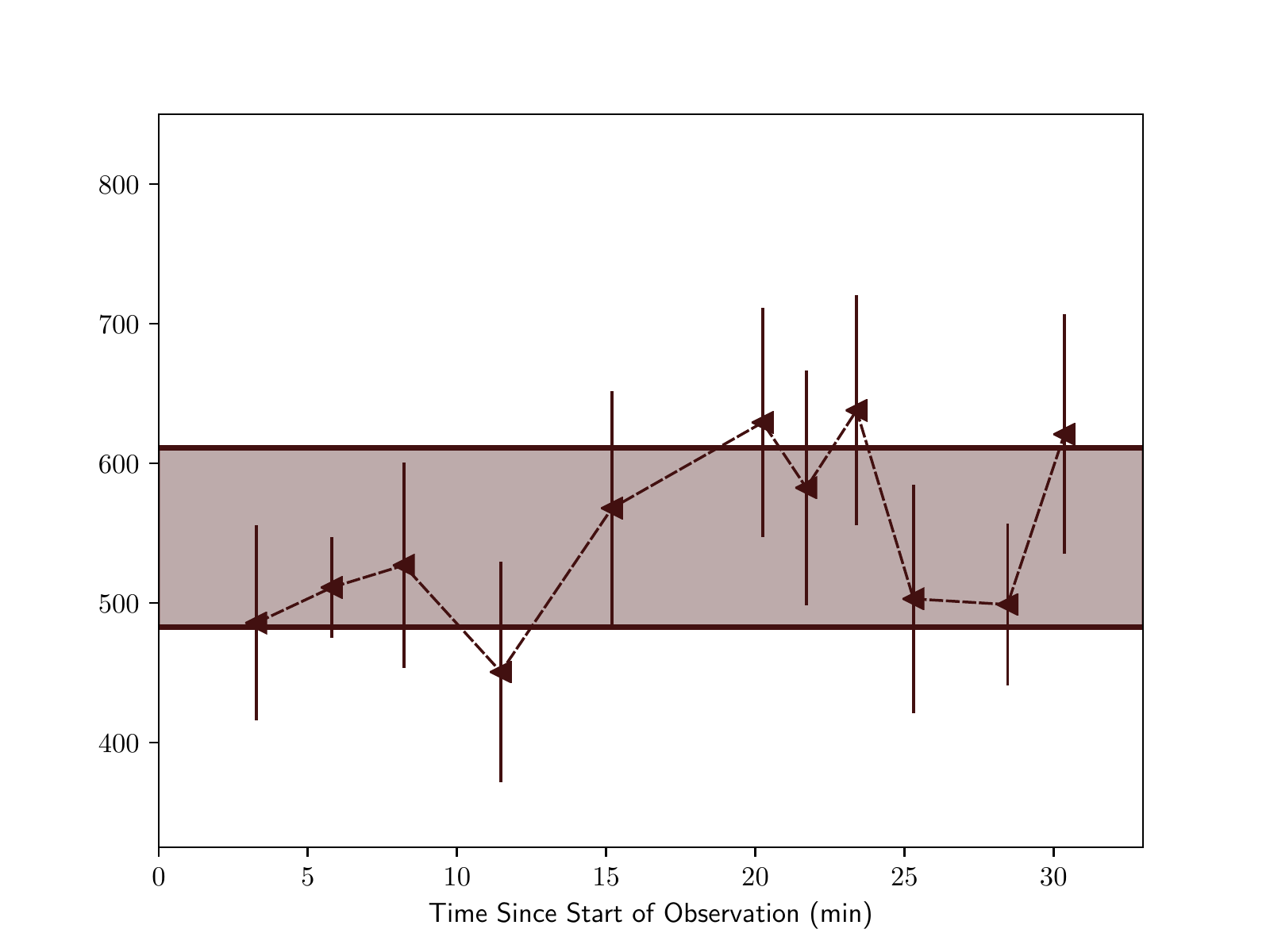}{0.427\textwidth}{}
          }       
\caption{The \longwave$\,$ peak brightness of \mbox{JW 566} (top left) along with 5 typical sources in the brightness range displayed by JW 566 throughout the 31 minute integration. The sources were selected from different locations around the \mbox{OMC 2/3} field. The light curve standard deviations are shown by the shaded regions in each plot. The average standard deviation of the non-varying sources (9.7\%) is overlaid on \mbox{JW 566}'s light curve.}
\label{fig:newvar-lightcurve}
\end{figure*}

Figure \ref{fig:newvar-lightcurve} shows a dramatic decay in brightness during the 31 minute integration.
The emission from \mbox{JW 566}
appears to already be in the dimming phase of the outburst, with an initial 
peak of \mbox{$773\mathrm{\:mJy\:beam}^{-1}$} that drops to 
\mbox{$400\mathrm{\:mJy\:beam}^{-1}$} by the end of the observation. The uncertainties for each measurement are calculated by measuring the square root average pixel variance in a $20\times20$ pixel box centered on the source. 

To confirm that this brightness decrease is significant, we also analyse the light 
curves of 5 non-varying unresolved sources with fluxes sampling the range of \mbox{JW 566} in this epoch. Subdividing the raw time stream was performed in the same way individually for each source as it was for \mbox{JW 566}. Data points with abnormally high variances due to their proximity to the edge of the map or the uncertainty in surrounding, large-scale structure have been discarded. 
All of these sources are consistent with a constant flux during the 31-minute observation. The average standard deviation of the non-varying sources is 9.7\%.  The lightcurve of \mbox{JW 566} is monotonically decreasing and has a standard deviation of 23.2\%, more than twice the average value of the non-varying sources. 
At \shortwave, the noise is too high to perform a similar analysis.

\section{Discussion} 
\label{sec:discuss}

Young stars are known to undergo large, short-lived (timescales of 
hours to days) outbursts detectable at millimetre and centimetre 
wavelengths 
\citep{bower2003,furuya2003,massi2006,salter2010,forbrich2017}. 
Flare studies tend to focus on these frequencies, leaving the sub-mm 
frequency space of JCMT largely unexplored. 

At a distance of \mbox{389 pc}\footnote{We adopt a distance of \mbox{389 pc} to \mbox{JW 566} 
due to its proximity with the Orion Nebula Cluster.  This is based on the analysis of \cite{kounkel2018}, who used Very Long Baseline Array data \citep{kounkel2017} and \textit{GAIA} DR2 astrometry \citep{Gaia2018}).}, the measured 
\longwave$\,$ flux ($466\mathrm{\:mJy\:beam}^{-1}$) corresponds to 
a radio luminosity of 
\mbox{$L_\nu= 8\times10^{19}\mathrm{\:erg\:s^{-1}\:Hz^{-1} }$}. A 
natural comparison point for this result is with the 2003 outburst 
event of the T Tauri star GMR-A in the Orion Nebula (\citealt{bower2003}; see also \citealt{furuya2003}). 
GMR-A had a 
radio luminosity of \mbox{$L_\nu= 3\times10^{19}\mathrm{\:erg\:s^{-1}\:Hz^{-1}}$} at \mbox{86 GHz} 
(assuming the same distance of \mbox{389 pc}), which makes the \mbox{JW 566} flare 
an order of magnitude brighter in terms of $\nu L_{\nu}$. \cite{salter2008} and \cite{massi2006} observed flares associated with the DQ Tau binary system at 115 GHz and the V773 Tau quadruplet at \mbox{90 GHz}, respectively, with radio luminosities of 
        $\sim6\times10^{19}\mathrm{\:erg\:s^{-1}\:Hz^{-1}}$.
If the flare associated with JW 566 follows the X-ray/radio 
luminosity correlation \citep[see, for example,][]{gudel2002}, then it
is 10 orders of magnitude brighter than a typical solar flare. It is plausible 
that this is the most luminous flare ever recorded in a 
young star. In the future, coordinated observations are 
required at \shortwave, \longwave, and other wavelengths to 
reveal the relationship between the fluxes at different 
energy regimes.

The OMC 2/3 field has been observed with SCUBA-2 for 10 hours since 
December 26, 2015, and this is the first significant flare event of 
its kind discovered in those data. In 
total, there are $\sim 600$ known (Spitzer identified; 
\citealt{megeath2012}) Class II (disk) objects present in the 
field of view. Therefore, the current detection rate of flare events 
of this magnitude is $1/(600\mathrm{\:stars}\times10\mathrm{\:hrs}) \approx 1-2\mathrm{\:yr}^{-1}\mathrm{\:star}^{-1}$.
It is likely that there is a luminosity function for submillimetre 
flares that scales as a power law, $N \propto L^{\beta}$, where 
$\beta < 0$. More and deeper observations of this field, including 
7 additional hours from our Transient survey by February 2020, will allow us to 
measure the flare rate over a wide range of luminosities. Additionally, we will be able to
perform a more complete search by detecting fainter, longer timescale events by co-adding subsets of the data.

The detection of a coronal flare at sub-mm wavelengths adds another 
source of uncertainty in the measurement of disk masses \citep[e.g.]
[]{pascucci2016}.  While sub-mm emission from most sources is 
produced by the thermal dust continuum emission within the 
protoplanetary disks, any unexpected emission from 
sources thought to be diskless should be tested to evaluate whether 
flaring may explain the emission.  Indeed, unresolved 1.3mm 
continuum emission from Prox Cen was initially interpreted as an indication of a 
candidate disk \citep{anglada2017} but later traced to a stellar flare 
\citep{macgregor2018}.

\subsection{Previous Observations of JW 566}

The  \mbox{JW 566} binary system ($0\farcs86$ projected separation) has a disk around at least one of the components \citep{megeath2012}.  \citet{daemgen2012} detected accretion around the K7 primary star but not the M1.5 secondary star.  High-resolution optical spectra of JW 566 are not available, so it is unknown whether one or both stars are spectroscopic binaries.  Interactions between the magnetospheres of close binaries are thought to excite coronal flares in DQ Tau \citep{salter2010} and perhaps other young stars.

Previous X-ray and radio observations demonstrate coronal flares 
from \mbox{JW 566}, as expected for young low-mass stars.
\cite{kounkel2014} classify the source as variable at 
both 4.5 and 7.5 GHz.
In addition, JW 566 is a known X-ray source 
\citep{gagne1995,garmire2000,feigelson2002,getman2005}, with variability on timescales of hours.
The JW 566 binary is one of the most luminous X-ray sources, with 
$L_X=10^{31}$ erg s$^{1}$, for its mass range in the COUP X-ray 
monitoring survey of the Orion Nebula \citep{getman2005}.  The 
extreme brightness is caused by a combination of saturated X-ray 
emission, with $\log L_X/L_{bol}=-3$, and large radii as measured by 
\citet{daemgen2012}.  The X-ray emission is harder than average but 
not extreme among the COUP sample.  

\begin{figure}
\centering
\includegraphics[width=0.5\textwidth]{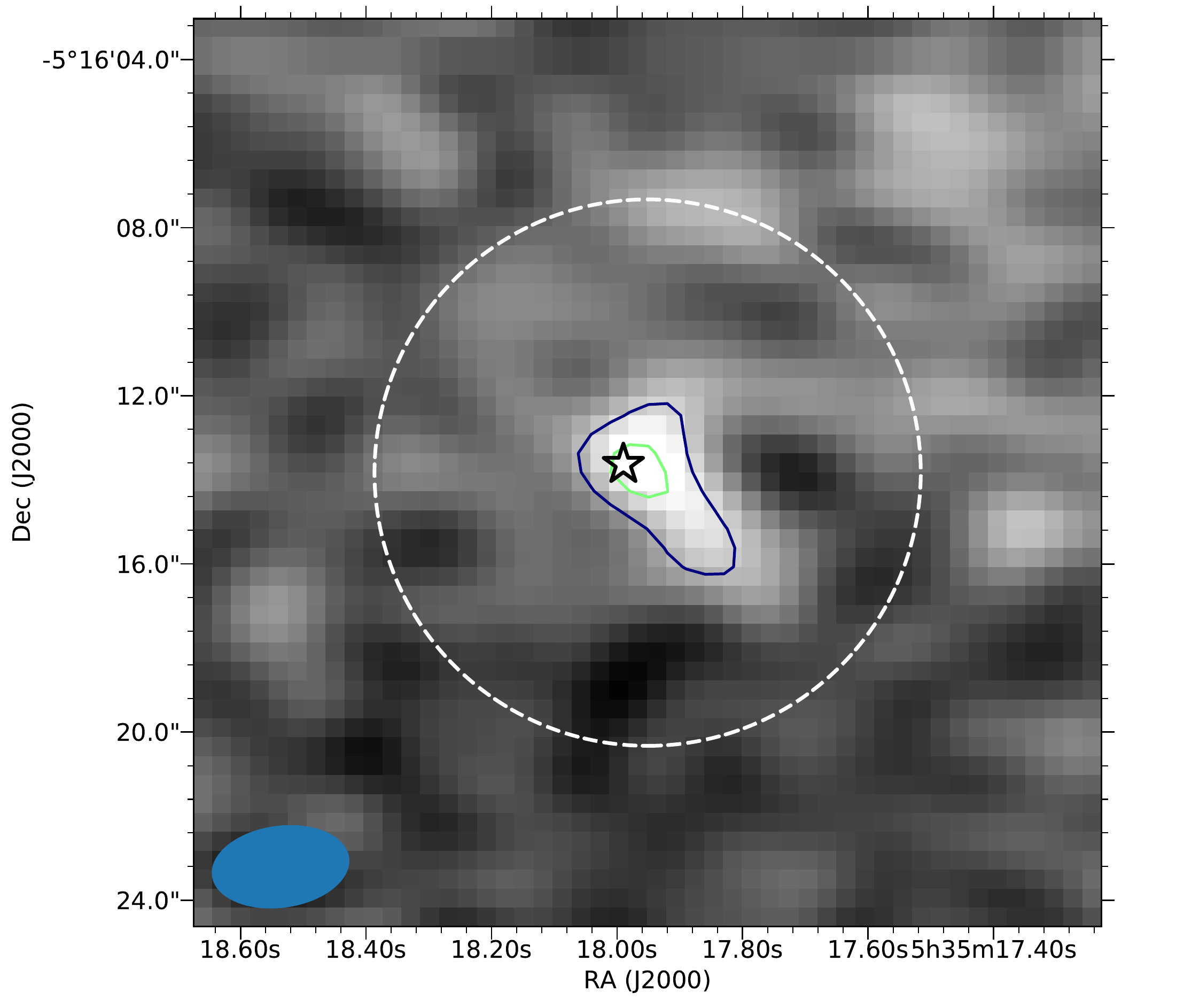}
\caption{3 mm continuum ALMA detection of \mbox{JW 566} (star symbol).  Contours are drawn at $3\sigma$ and $5\sigma$. The dashed circle indicates the size of the SCUBA-2 \longwave$\,$ beam FWHM. The beam is displayed in the lower left.}
\label{fig:ALMA}
\end{figure}

The source is detected in a \mbox{3
$\mathrm{\:mm}$} continuum image obtained by ALMA on 2015-12-26 
(\citealt{hacar2018}; see Figure \ref{fig:ALMA}). The 3 mm flux is 
measured to be \mbox{$0.6 \mathrm{\:mJy}$}, yielding a SNR of 5.7. This 
presumably-quiescent flux is a factor of $8\times10^{2}$ fainter than the 
\longwave$\,$ continuum measurement of the flare, assuming a spectral index of 1.  If the 3 mm flux is produced by the disk, a spectral index of $\sim2.3$ \citep[e.g.][]{ricci2010} would lead to an 850 $\mu$m flux of $\sim10\mathrm{\:mJy\:beam}^{-1}$, very close to our current detection limit in the co-added image and within the flux range expected for disks in nearby star-forming regions \citep[e.g.][]{ansdell2016,pascucci2016}.

Significant variability has not been detected at wavelengths other 
than millimetres and the X-rays, including in the optical, 
$JHK_S$ \citep{tsujimoto2003,skrutskie2006,ali1995}, at 3.6 and 4.5 
$\mu$m \citep{moralescaldero2011}, or in the far-IR 
\citep{billot2012}. 
Unfortunately, we do not 
know of other available data at optical, infrared, or radio 
wavelengths at the time of this observation.

\subsection{The Nature of JW 566's Flare}
\label{sec:discussFlare}

Based on the 
measured source fluxes at \shortwave$\,$ (666~GHz, \mbox{$f_{666} = 500\pm107\mathrm{\:mJy\:beam}^{-1}$}) and \longwave$\,$ (352~GHz, \mbox{$f_{352} = 466\pm46.6\mathrm{\:mJy
\:beam}^{-1}$}) we calculate a spectral index

\begin{equation}
\alpha = \frac{\left(\mathrm{log}(f_{666})\pm\frac{\Delta f_{666}}{f_{666}\mathrm{ln}(10)}\right) - \left(\mathrm{log}(f_{352})\pm\frac{\Delta f_{352}}{f_{352}\mathrm{ln}(10)}\right)}{\mathrm{log}(666\mathrm{\:GHz}) - \mathrm{log}(352\mathrm{\:GHz})},
\end{equation}

\noindent where $\Delta f_{\nu}$ represents the uncertainty in the fluxes, of \mbox{$\alpha = 0.11\pm0.49$} over these wavelengths.
While this value is consistent with non-thermal emission\footnote{Assuming a blackbody thermal spectrum, the temperature would need to be 4.9~K to reproduce such a flat spectral index.}, the spectral index itself is not sufficient
to discriminate the emission mechanism.

The brightness temperature, $T_{b}$ can be approximated by

\begin{equation}
T_{b} \sim \left(\frac{1}{2k_{B}}\right)\times \lambda^{2}\Delta S\times\left(\frac{D}{c\Delta t}\right)^{2}
\end{equation}

\noindent where $k_{B}$ is Boltzmann's constant, $\Delta S$ 
is the change in flux over time $\Delta t$, $\lambda$ is the 
wavelength, $D$ 
is the distance to the source (\mbox{389 pc}), and $c$ is the speed of 
light. 
The change in \longwave$\,$ flux ($\Delta S$) and its corresponding 
time frame ($\Delta t$) are derived from the light curve 
presented in Figure \ref{fig:newvar-lightcurve} with values 
of \mbox{$373\mathrm{\:mJy}$} \mbox{(mJy = mJy beam$^{-1}$, 
for point sources)} and \mbox{$1661 \mathrm{\:sec}$}, 
respectively. This results in a brightness temperature of 
\mbox{$T_{b} \sim 6\times10^{4}\mathrm{\:K}$} and a light crossing
distance of 3.3~AU. 
Constraining the angular scale to a stellar 
radius of \mbox{$R_{\star} = 2.5\mathrm{\:R}_{\odot}$} \citep[the 
average radius of the two components in the JW 566 binary system; ][]{daemgen2012} rather than $c\Delta t$, 
results in an estimate of the upper range of the brightness temperature of \mbox{$T_{b} \sim 5\times10^{9}
\mathrm{\:K}$}, though the origin of the flare could indeed be generated by a region smaller in scale than 
\mbox{$R_{\star} = 2.5\mathrm{\:R}_{\odot}$}.
With a projected separation of \mbox{335 AU},  it is very unlikely 
that an inter-binary interaction is occurring among the known 
components.  These calculations strongly favour non-thermal emission. 
The most likely scenario is that the flare was caused by gyrosynchrotron/synchrotron radiation emitted by a reconnection event 
in the strong magnetic fields present in the corona of these 
young stars \citep[see, e.g., ][]{salter2010}. This 
magnetic reconnection briefly energizes non-thermal 
particles which appears as a flare. The detection of such an 
event at \longwave$\,$ suggests a very high energy acceleration 
of electrons. Polarimetry data is required to separately constrain the contributions of gyrosynchrotron and 
synchrotron emission; such data, however, are not available for this event.

\section{Summary}
\label{sec:summary}

In this paper, we presented \shortwave$\,$ and \longwave$\,$ 
SCUBA-2 observations of a bright flare associated with the T Tauri 
binary system \mbox{JW 566} (R.A., Decl. = 5:35:17.94,$-5$:16:11, J2000) obtained by the JCMT Transient Survey on 2016-11-26 (UT). 
The flare is measured to have a flux of \mbox{$466
\pm47\mathrm{\:mJy\:beam}^{-1}$} at \longwave$\,$ and \mbox{$500
\pm107\mathrm{\:mJy\:beam}^{-1}$} at \shortwave$\,$, averaged over the observation (see Sections 
\ref{sec:850detection} and \ref{sec:450detection}). We subdivided the 31 minute integration into nine intervals based on when the telescope scanned over the source and found a monotonic 
decrease of $337\mathrm{\:mJy\:beam}^{-1}$ over $1661\mathrm{\:sec}$ 
(see Section \ref{lightcurve}). Constraining the size scale of the 
flare origin to the light crossing time of our observation and to the stellar radius of one of the binary components 
results in a range of brightness temperatures between \mbox{$T_{b}\sim6\times10^{4}\mathrm{\:K}$} and  \mbox{$T_{b} 
\sim5\times10^{9}\mathrm{\:K}$} (see Section 
\ref{sec:discussFlare}). The flat spectral index, the short variability timescale, and a large $T_{b}$ strongly indicate the flare is a result of  
gyrosynchrotron/synchrotron emission, likely caused by a magnetic reconnection event. 

The true timescale of the flare remains unknown, as there are no data available at other wavelengths during the time of our observation. The 
JCMT Transient Survey will continue through January 2020, increasing the number of observations of the OMC 2/3 field by a factor of $
\sim1.7$. Therefore, it is plausible another burst of a similar magnitude will be detected. With new variable source detection methods 
Lalchand et al. (in prep.), we will be able to identify future events within $\sim24\mathrm{\:hours}$ of the data being taken by the telescope 
in order to perform follow-up observations.  


\acknowledgments

The authors wish to 
extend their gratitude to Dr. David Berry for useful discussions 
regarding the \shortwave$\,$ data reduction, to Dr. Helen Kirk for 
assistance with the JCMT GBS DR3 observations, to Dr. Jenny Hatchell who provided 
comments which strengthened this work, and to the
anonymous referee for insightful comments.

\vspace{2mm}

DJ is supported by NRC Canada and an NSERC Discovery Grant. GJH is 
supported by general grants 11473005 and 11773002 awarded by the 
National Science Foundation of China. JEL is supported by the Basic 
Science Research Program through the National Research Foundation of 
Korea (grant No. NRF-2018R1A2B6003423) and the Korea Astronomy and 
Space Science Institute under the R\&D program supervised by the 
Ministry of Science, ICT and Future Planning. AH thanks the support of 
the Netherlands Organisation for Scientific Research (NWO)
under the VENI project 639.041.644. This paper makes use of the 
following ALMA data: ADS/JAO.ALMA\#2015.1.00669.S. 

\vspace{2mm}

The authors wish to recognise and acknowledge the very significant 
cultural role and reverence that the summit of
Maunakea has always had within the indigenous Hawaiian community. 
We are most fortunate to have the opportunity
to conduct observations from this mountain. 
The James Clerk Maxwell Telescope is operated by the East Asian 
Observatory on behalf of The National Astronomical Observatory of 
Japan; Academia Sinica Institute of Astronomy and Astrophysics; the 
Korea Astronomy and Space Science Institute; the Operation, 
Maintenance and Upgrading Fund for Astronomical Telescopes and 
Facility Instruments, budgeted from the Ministry of Finance (MOF) 
of China and administrated by the Chinese Academy of Sciences 
(CAS), as well as the National Key R\&D Program of China (No. 
2017YFA0402700). Additional funding support is provided by the 
Science and Technology Facilities Council of the United Kingdom and 
participating universities in the United Kingdom and Canada. 
Additional funds for the construction of SCUBA-2 were provided by 
the Canada Foundation for Innovation. This research used the 
facilities of the Canadian Astronomy
Data Centre operated by the National Research Council of
Canada with the support of the Canadian Space Agency. This research has made use of the SIMBAD database,
operated at CDS, Strasbourg, France \citep{simbad}. 

\facilities{JCMT, ALMA}

\software{astropy \citep{astropy},  
          matplotlib \citep{matplotlib},
          aplpy \citep{aplpy}
          starlink \citep{currie2014}, 
          }

\bibliography{OMC23var}

\end{document}